RESEARCH ARTICLE

# Entropy-Based Financial Asset Pricing

**Mihály Ormos\*, Dávid Zibriczky**

Department of Finance, Budapest University of Technology and Economics, Magyar tudósok krt. 2., 1117, Budapest, Hungary

\*ormos@finance.bme.hu

## Abstract

We investigate entropy as a financial risk measure. Entropy explains the equity premium of securities and portfolios in a simpler way and, at the same time, with higher explanatory power than the beta parameter of the capital asset pricing model. For asset pricing we define the continuous entropy as an alternative measure of risk. Our results show that entropy decreases in the function of the number of securities involved in a portfolio in a similar way to the standard deviation, and that efficient portfolios are situated on a hyperbola in the expected return – entropy system. For empirical investigation we use daily returns of 150 randomly selected securities for a period of 27 years. Our regression results show that entropy has a higher explanatory power for the expected return than the capital asset pricing model beta. Furthermore we show the time varying behavior of the beta along with entropy.

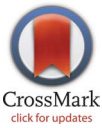

**Citation:** Ormos M, Zibriczky D (2014) Entropy-Based Financial Asset Pricing. PLoS ONE 9(12): e115742. doi:10.1371/journal.pone.0115742

**Editor:** Giampiero Favato, Kingston University London, United Kingdom

**Received:** August 8, 2014

**Accepted:** November 26, 2014

**Published:** December 29, 2014

**Data Availability:** The authors confirm that all data underlying the findings are fully available without restriction. Data are from the Center for Research in Security Prices (http://www.crsp.com/). Access to CRSP data requires a subscription. For subscription in formation please contact subscriptions@crsp.chicagobooth.edu.

**Funding:** The authors have no support or funding to report.

**Competing Interests:** The authors have declared that no competing interests exist.

## Introduction

We build an equilibrium capital asset pricing model by applying a novel risk measure, the entropy. Entropy characterizes the uncertainty or measures the dispersion of a random variable. In our particular case, it characterizes the uncertainty of stock and portfolio returns. In modern Markowitz [1] portfolio theory and equilibrium asset pricing models [2] we apply linear regressions. This methodology supposes that the returns are stationary and normally distributed; however, this is not actually the case [3]. Entropy, on the other hand, does not have this kind of boundary condition. The main goal of this paper is to apply entropy as a novel risk measure. As a starting point even the density function itself has to be estimated. In the traditional asset pricing model there is equilibrium between expected return the beta parameter, which is the covariance–variance ratio between the market portfolio and the investigated investment opportunity. If the random variable is normally distributed then the entropy follows its standard deviation; thus in the ideal case there is no difference between the two risk measures. However; our results show that there is a significant difference between the standard deviation, or beta, and the entropy of a given security or portfolio. In this paper we show that entropy offers an ideal alternative for capturing the risk of an investment opportunity. If we explain the return of a wide sample of securities and portfolios with different risk measures then on an ordinary least squares (OLS) regression setting the explanatory power is much higher in the case of the entropy measure of risk than in the case of the traditional measures, both in-sample and out-of-sample. We show that entropy reduction in line with diversification behaves similarly to standard deviation; however at the same time it captures a beta-like systematic risk of single securities or non-efficient portfolios as well. For well-diversified portfolios the explanatory power of entropy is 1.5 times higher than that of the capital asset pricing model (CAPM) beta.

We also test and compare entropy with standard risk measures for market circumstances that are increasing and decreasing, and find that the explanatory power of entropy is significantly higher in a bullish market, but lower for a bearish market. Our results for bullish and bearish regimes show that the different risk measures behave similarly in terms of the positive and negative relationship between risk and return. This behavior underlines the fact that the entropy-based risk measure can give contradictory results in the same way as traditional risk estimations in upward and downward regimes.

We also compare the entropy-based risk measures with the CAPM beta in and out of sample, which gives information on the predictive power of the different methods. As the CAPM beta measures the systematic risk only, while entropy based risk measures and the standard deviation captures the total risk of the investment our results are shocking, that entropy gives almost twice as high an average explanatory power as the beta, with an average of 40% less standard deviation. A further contribution of the paper is that we introduce a simple method to estimate the entropy of a security or portfolio return.

## Data

In our empirical analysis we apply daily returns from the Center for Research in Security Prices (CRSP) database for the period from 1985 to the end of 2011. We randomly select 150 securities from the S&P500 index components that are available for the full period. The market return is the CRSP value-weighted index return premium above the risk-free rate. The index tracks the return of the New York Stock Exchange (NYSE), the American Stock Exchange (AMEX) and NASDAQ stocks. The risk-free rate is the return of the one-month Treasury bill from the CRSP. We use daily returns because they are not normally distributed (see S1 Table). Erdős and Ormos (2009) [3] and Erdős et al. (2011) [4] describe the main difficulties of modeling asset prices with non-normal returns. The daily return calculation enables us to compare different risk measures.

## Methodology

Entropy is a mathematically-defined quantity that is generally used for characterizing the probability of outcomes in a system that is undergoing a process. It was originally introduced in thermodynamics by Rudolf Clausius [5] to measure the ratio of transferred heat through a reversible process in an isolated system. In statistical mechanics the interpretation of entropy is the measure of uncertainty about the system that remains after observing its macroscopic properties (pressure, temperature or volume). The application of entropy in this perspective was introduced by Ludwig Boltzmann [6]. He defined the configuration entropy as the diversity of specific ways in which the components of the system may be arranged. He found a strong relationship between the thermodynamic and the statistical aspects of entropy: the formulae for thermodynamic entropy and configuration entropy only differ in the so-called Boltzmann constant. There is an important application of entropy in information theory as well, and this is often called Shannon [7] entropy. The information provider system operates as a stochastic cybernetic system, in which the message can be considered as a random variable. The entropy quantifies the expected value of the information in a message or, in other words, the amount of information that is missing before the message is received. The more unpredictable (uncertain) the message that is provided by the system, the greater the expected value of the information contained in the message. Consequently, greater uncertainty in the messages of the system means higher entropy. Because the entropy equals the amount of expected information in a message, it measures the maximum compression ratio that can be applied without losing information.

In financial applications, Philippatos and Wilson [8] find that entropy is more general and has some advantages over standard deviation; in their paper they compare the behaviors of standard deviation and entropy in portfolio management. Kirchner and Zunckel [9] argue that in financial economics entropy is a better tool for capturing the reduction of risk by diversification; however, in their study they suppose that the assets are Gaussian. Dionisio et al. [10] argue that entropy observes the effect of diversification and is a more general measure of uncertainty than variance, since it uses more information about the probability distribution. The mutual information and the conditional entropy perform well when compared with the systematic risk and the specific risk estimated through the linear equilibrium model. Regarding the predictability of stock market returns, Maasoumi and Racine [11] point out that entropy has several desirable properties and is capable of efficiently capturing nonlinear dependencies in return time series. Nawrocki and Harding [12] propose applying state-value weighted entropy as a measure of investment risk; however, they are dealing with the discrete case.

All the above academic papers recognize that entropy could be a good measure of risk; however, it seems to be difficult to use this measure. Our main motivation is to show that an entropy-based risk measure is, on the one hand, more precise, and, on the other hand, no more complicated to use than variance equilibrium models.

## Discrete entropy function

Entropy functions can be divided into two main types, discrete and differential entropy functions.

Let $X^*$ be a discrete random variable. The possible outcomes of this variable are denoted by $o_1, o_2, .., o_n$, and the corresponding probabilities by $p_i = \Pr(X^* = o_i)$, $p_i \geq 0$ and $\sum_{i=1}^{n} p_i = 1$. The generalized discrete entropy function [13] for the variable $X^*$ is defined as:

$$H_\alpha(X^*) = \frac{1}{1-\alpha} \log \left( \sum_{i=1}^{n} p_i^\alpha \right), \tag{1}$$

where $\alpha$ is the order of entropy, $\alpha \geq 0$ and $\alpha \neq 1$, and the base of the logarithm is 2. The order of entropy expresses the weight taken into account in each outcome; if the order of entropy is lower, the more likely outcomes are underweighted, and vice versa. The most widely used orders are $\alpha = 1$ and $\alpha = 2$.

$\alpha = 1$ is a special case of generalized entropy. However the substitution of $\alpha = 1$ into (1) results in a division by zero. It can be shown, using l'Hôpital's rule for the limit of $\alpha = 1$, that $H_\alpha$ converges to the Shannon entropy:

$$H_1(X^*) = - \sum_{i=1}^{n} p_i \log(p_i) \tag{2}$$

The case of $\alpha = 2$ is called collision entropy and similarly to the literature we refer to this special case as "Rényi entropy" further in the paper:

$$H_2(X^*) = - \log \left( \sum_{i=1}^{n} p_i^2 \right) \tag{3}$$

$H_\alpha(X)$ is a non-increasing function in $\alpha$, and both entropy measures are greater than zero provided that there is a finite number of possible outcomes:

$$0 < H_2(X^*) \leq H_1(X^*) \tag{4}$$

## Differential entropy function

Let $X$ be a continuous random variable taking values from $\mathbb{R}$ with a probability density function $f(x)$. Analogously to (1), the continuous entropy is defined as:

$$H_\alpha(X) = \frac{1}{1-\alpha}\ln \int f(x)^\alpha dx \qquad (5)$$

One can see that the bases of the logarithms in (1) and (5) are different. Although the entropy depends on the base, it can be shown that the value of the entropy changes only by a constant coefficient for different bases. We use the natural logarithm for all differential entropy functions. The formulas for the special cases ($\alpha=1$ and $\alpha=2$) are the following:

$$H_1(X) = -\int f(x)\ln f(x)dx \qquad (6)$$

$$H_2(X) = -\ln \int f(x)^2 dx \qquad (7)$$

An important difference between discrete and continuous entropy is that while discrete entropy takes only non-negative values, continuous entropy can also take negative values:

$$H_\alpha(X) \in \mathbb{R} \qquad (8)$$

In practice, standard risk measures like the CAPM beta or standard deviation are calculated on daily or monthly return data. We also follow this practice, and use a formula that is able to capture risk using this kind of data. Since the return on securities can take values from a continuous codomain, we primarily focus on the differential entropy function. However, by grouping return values into bins the discrete entropy function may also be used; this solution is outside the scope of this paper.

## Entropy estimation

For the estimation of differential entropy, the probability density function of the return values needs to be estimated. Let $x_1, x_2, ..., x_n$ be the observations of the continuous random variable $X$, and $H_{\alpha,n}(X)$ the sample-based estimation of $H_\alpha(X)$. The plug-in estimations of entropy are calculated on the basis of the density function estimation. The probability density function $f(x)$ is estimated by $f_n(x)$, the integral estimate of entropy, in the following way:

$$H_{\alpha,n}(X) = \frac{1}{1-\alpha}\ln \int_{A_n} f_n(x)^\alpha dx, \qquad (9)$$

where $A_n$ is the range of integration, which may exclude small and tail values of $f_n(x)$. We propose to select $A_n = (\min(x), \max(x))$.

## Histogram

One of the simplest methods of density estimation is the histogram-based density estimation. Let $b_n = (\max(x), \min(x))$ be the range of sample values; partition the range into $k$ bins of equal width and denote the cutting points by $t_j$. The width of a bin is constant: $h = \frac{b_n}{k} = t_{j+1} - t_j$. The density function is estimated by using the following formula:

$$f_n(x) = \frac{v_j}{nh},\tag{10}$$

if $x(t_j, t_{j+1})$, where $v_j$ is the number of data points falling in the $j^{\text{th}}$ bin.

Based on the properties of the histogram, a simpler non plug-in estimation formula can be deduced for Shannon and Rényi entropy using (6), (7), (9) and (10):

$$H_{1,n}(X) = \frac{1}{n}\sum_{j=1}^{k} v_j \ln\left(\frac{v_j}{nh}\right)\tag{11}$$

$$H_{2,n}(X) = -\ln \sum_{j=1}^{k} h\left(\frac{v_j}{nh}\right)^2\tag{12}$$

The parameter of this method is the number of equal width bins ($k$). However, there are several methods for choosing this parameter (e.g. the square root choice, Scott's normal reference rule [14], or the Freedman-Diaconis rule [15]); the detailed descriptions of these are outside the scope of this paper.

## Kernel density estimation

The kernel-based density estimation is another commonly used method. It applies the following formula:

$$f_n(x) = \frac{1}{nh}\sum_{i=1}^{n} K\left(\frac{x - x_i}{h}\right),\tag{13}$$

where $K()$ is the kernel function, and $h$ is the bandwidth parameter. There are several kernel functions that can be used (see S2 Table); for practical reasons (computational time), we propose using the indicator-based Epanechnikov kernel function:

$$K(z) = \frac{3}{4}\left(1 - z^2\right)\mathrm{I}_{\{|z| \leq 1\}},\tag{14}$$

where I is the indicator function.

Härdle [16] shows that the choice of the kernel function is only of secondary importance, so the focus is rather on the right choice of bandwidth ($h$). One of the most widely used simple formulas for the estimation of $h$ is Silverman's rule of thumb [17]:

 

$$\hat{h}_{rot} = 1.06 \, \min \left\{ \sqrt{\frac{1}{n-1} \sum_{i=1}^{n} (x_i - \bar{x})^2}, \frac{IQR(x)}{1.34} \right\} n^{-\frac{1}{5}}, \tag{15}$$

where $IQR(x)$ is the interquartile range of $x$.

As the formula assumes a normal distribution for $X$ it gives an approximation for optimal bandwidth; despite this, Silverman's rule of thumb can be used for a good initial value for more sophisticated optimization methods [18].

**Sample spacing estimation**

Let $x_{n,1} \le x_{n,2} \le \ldots \le x_{n,n}$ be the corresponding order of $x_1, x_2, \ldots, x_n$, assuming that this is a sample of i.i.d. real-valued random variables. $x_{n,i+m} - x_{n,i}$ is called a spacing of order $m$ ($1 \le i < i+m < n$). The simple sample spacing density estimate is the following [19]:

$$f_n(x) = \frac{m}{n} \frac{1}{x_{n,im} - x_{n,(i-1)m}}, \tag{16}$$

if $x[x_{n,(i-1)m}, x_{n,im})$.

Wachowiak et al. [20] introduced another variation of the sample spacing density estimation, called the Correa estimator:

$$f_n(x) = \frac{1}{n} \frac{\sum_{j=i-m/2}^{i+m/2} (x_j - \bar{x}_i)(j-i)}{\sum_{j=i-m/2}^{i+m/2} (x_j - \bar{x}_i)^2}, \tag{17}$$

if $i: x \in [x_{n,i}, x_{n,i+1})$; $\bar{x}_i = \frac{1}{m+1} \sum_{j=i-m/2}^{i+m/2} x_j$, and $1 \le j \le n$.

The parameter for sample spacing methods is the fixed order $m$. For practical reasons (e.g. different sizes of samples) we suggest using $m_n$, which depends on the size of the sample and is calculated by the following formula:

$$m_n = \lceil \frac{n}{k} \rceil, \tag{18}$$

where $k$ is the number of bins, and the braces indicate the ceiling function.

Beirlant et al. [19] overview several additional entropy estimation methods, such as resubstitution, splitting-data and cross-validation; however, our paper focuses on the applications that are used most often.

## Risk estimation

Let the following be a given set of data:

$$D : \{S, R, R_M, R_F\} \tag{19}$$

The elements are the set of securities $S:\{S_1, S_2,...,S_l\}$, with the corresponding observations being $R:\{R_1, R_2,...,R_l\}$, where $R_i=(r_{i1}, r_{i2},...,r_{in})$. The observation for the market return is $R_M=(r_{M1}, r_{M2},...,r_{Mn})$, and the observation for the risk free return is $R_F=(r_{F1}, r_{F2},...,r_{Fn})$ where $l$ is the number of securities and $n$ is the number of samples. Let us recall that the main goal of this paper is to apply entropy as a novel risk measure. In order to handle the risk measure uniformly, we introduce $\kappa$ as a unified property for securities. Let $\kappa(S_i)$ be the risk estimate for the security $i$.

In the economic literature the most widely used risk measures are the standard deviation and the CAPM beta. Let us denote these by $\kappa_\sigma$ and $\kappa_\beta$, respectively. The estimation of these risk measures for the security $i$ is the following:

$$\hat{\kappa}_\sigma(S_i) = \sigma(R_i - R_F) \qquad (20)$$

and

$$\hat{\kappa}_\beta(S_i) = \beta_i = \frac{\text{cov}(R_i - R_F, R_M - R_F)}{\sigma^2(R_M - R_F)}, \qquad (21)$$

where $\beta$ is the CAPM beta, $\text{cov}()$ is the covariance of the arguments and $\sigma$ is the standard deviation.

Our hypothesis is that uncertainty about the observation values can be interpreted as a risk of the security, and for this reason we apply entropy as a risk measure. Because the differential entropy function can also take negative values, for better interpretability we apply the exponential function to the entropy, and we define the entropy-based risk measure by the following formula:

$$\hat{\kappa}_H(S_i) = e^{H_n(R_i - R_F)} \qquad (22)$$

One can see that $\kappa_H$ takes values from the non-negative real numbers, $\kappa_H \in [0, +\infty)$.

## Explanatory and predictive power

In order to compare the efficiency of the risk estimation methods, we introduce two basic evaluation approaches, the measurement of in-sample explanatory power and the measurement of out-of-sample predictive power.

### In-sample

Let $V$ be a target variable, with sample $v=(v_1, v_2,...,v_l)$, and let $U$ be a single explanatory variable with sample $u=(u_1, u_2,...,u_l)$. To estimate the explanatory power of the variable $U$ for the variable $V$, we use the following method. The linear relationship between the two variables can be described using the linear regression model: $V = a_0 + a_1 U + \varepsilon$.

The parameters of the model ($a_0$ and $a_1$) are estimated by ordinary least squares (OLS), and the estimation for the target value is the following: $\hat{v}_i = \hat{a}_0 + \hat{a}_1 u_i$ where $\hat{a}_0$ and $\hat{a}_1$ are the estimations of $a_0$ and $a_1$, respectively. One of the most commonly applied estimations of the explanatory power is the $R^2$ (goodness of fit, or coefficient of determination) of the linear regression:

$$R^2(v,u) = 1 - \frac{\sum_{i=1}^{n} \left( v_i - (\hat{a}_0 + \hat{a}_1 u_i) \right)^2}{\sum_{i=1}^{n} (v_i - \bar{v})^2} \tag{23}$$

We are curious as to how efficiently the different risk measures describe the expected return of a security, and we denote this measure by $\eta(\kappa)$. Let the explanatory variable $U$ be the risk measure of the securities, where the sample is:

$$u_\kappa = (\hat{\kappa}(S_1), \hat{\kappa}(S_2), \dots, \hat{\kappa}(S_l)), \tag{24}$$

and the target variable $T$ is the expected risk premium of the securities, where the sample is:

$$v_\mu = (E[R_1 - R_F], E[R_2 - R_F], \dots, E[R_l - R_F]), \tag{25}$$

where $\kappa$ is the unified risk measure function, and $E[]$ is the expected value of the argument. We define the estimation of the in-sample explanatory power (efficiency) as the $R^2$ of the previously defined variables (24) and (25):

$$\hat{\eta}(\kappa) = R^2(v_\mu, u_\kappa) \tag{26}$$

**Out of sample**

Let us create a split of samples for a given D:$\{S, R, R_M, R_F\}$ data set (19):

$$D^I : \left\{ S^I, R^I, R_M^I, R_F^I \right\}, \quad D^O : \left\{ S^O, R^O, R_M^O, R_F^O \right\}, \tag{27}$$

where the corresponding samples for the securities are $R^I : \left\{ R_1^I, R_2^I, \dots, R_l^I \right\}$, $R_i^I = (r_{i1}, r_{i2}, \dots, r_{im})$ and $R^O : \left\{ R_1^O, R_2^O, \dots, R_l^O \right\}$, $R_i^O = (r_{i(m+1)}, r_{i(m+2)}, \dots, r_{i(m+p)})$, the split for market returns is $R_M^I = (r_{M1}, r_{M2}, \dots, r_{Mm})$ and $R_M^O = (r_{M(m+1)}, r_{M(m+2)}, \dots, r_{M(m+p)})$, and the split for the risk free rates is $R_F^I = (r_{F1}, r_{F2}, \dots, r_{F(m+p)})$ and $R_F^O = (r_{F(m+1)}, r_{F(m+2)}, \dots, r_{F(m+p)})$, where $|S^I| = |S^O|$, $|R_i^I| = m$, $|R_i^O| = p$, $(1 \leq i \leq l)$, and $m+p = n$.

The explanatory values contain the risk estimates for the set of securities based on the data set $D^I$:

$$u_\kappa^I = (\hat{\kappa}(S_1^I), \hat{\kappa}(S_2^I), \dots, \hat{\kappa}(S_l^I)), \tag{28}$$

and the target values are the expected risk premium of the securities based on $D^O$:

$$v_\mu^O = \left( E\left[ R_1^O - R_F^O \right], E\left[ R_2^O - R_F^O \right], \dots, E\left[ R_l^O - R_F^O \right] \right) \tag{29}$$

Based on (26), (28) and (29), the estimation of the out-of-sample explanatory (predictive) power is the following:

$$\hat{\eta}_O(\kappa) = R^2\left(v_\mu^O, u_\kappa^I\right) \qquad (30)$$

Both in- and out of sample we test whether the difference between the explanatory power of the investigated risk measures (standard deviation, CAPM beta, Shannon- and Rényi entropy) are significant by applying bootstrapping method. In our bootstrap iteration we remove 25 random stocks from the investigated 150 ones and measure the $R^2$s of the four different models. We apply 1000 iterations to approximate the distribution of $R^2$ values on random selection, and we test the equality of means of $R^2$s by applying *t-test* on the generated samples.

## Results and Discussion

We present the empirical results in four parts. First, we show how the entropy behaves in the function of securities involved into the portfolio. Second, we present the long-term explanatory power of the investigated models. Third we examine and compare the performance of different risk measures in in upward and downward market trends. Fourth we apply the different risk parameters to predict future returns, thus we test the out of sample explanatory power of the well-known risk parameters and compare their efficiency to the entropy based risk measures.

### Characterizing the diversification effect

We investigate whether entropy is able to measure the reduction of risk by diversification. We generate 10 million random equally-weighted portfolios with different numbers of securities involved (at most 100,000 for each size), based on the 150 randomly selected securities from the S&P500. The risk of portfolios is estimated by standard deviation, and by the Shannon and Rényi entropies using risk premiums for the full period. Because the CAPM beta measures the systematic risk only, we exclude it from the investigation of risk reduction. Both types of entropy functions are calculated by the histogram-based density function estimation, with 175 bins for the Shannon entropy and 50 bins for the Rényi entropy. (We tested the histogram, sample spacing and kernel density estimation methods, and the histogram-based method proved to be the most efficient in terms of explanatory and predictive power and simplicity. See our results in S3 Table.)

Fig. 1 shows the diversification effects that are characterized by the entropic risk measures and by the standard deviation. For 10 random securities involved in the portfolio, approximately 40% of risk reduction can be achieved compared to a single random security, based on all of the three risk estimators under investigation.

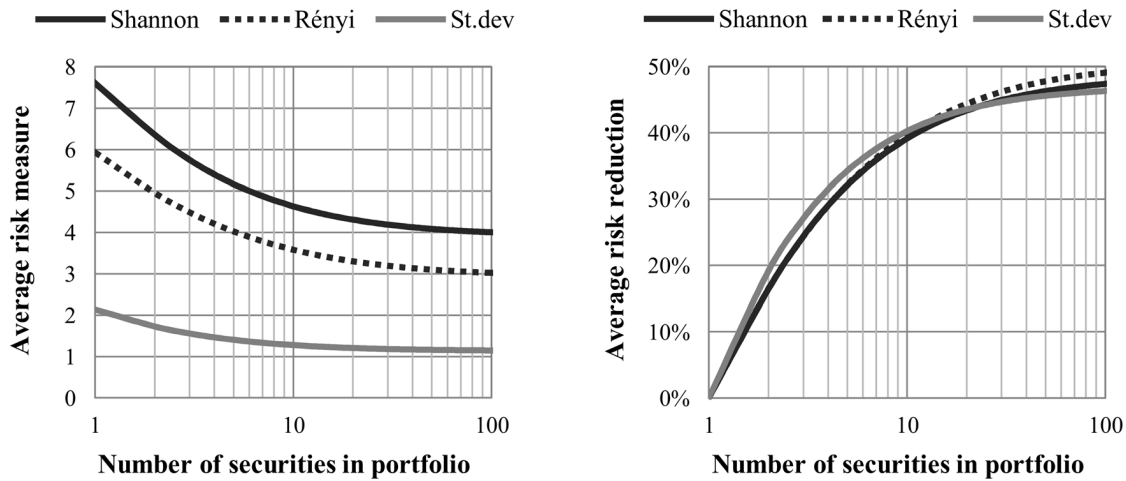

**Fig. 1. Average value of risk and risk reduction vs. number of securities in portfolio.** *Note*: We generate 10 million random equally weighted portfolios with different number of securities involved (at most 100,000 for each size) based on 150 randomly selected securities from S&P500. The risk of portfolios is estimated by standard deviation (gray continuous curve), Shannon- (black continuous curve) and Rényi entropy (black dashed curve) in the period from 1985 to the end of 2011. Both types of entropy functions are calculated by histogram based density function estimation. The left chart shows the average risk estimates for each portfolio size, and the right chart shows the risk reduction compared to an average risk of single security portfolio.

Fig. 1 suggests that entropy shows behavior that is similar to but not the same as standard deviation, so it can serve as a good measure of risk. We also investigate how the different portfolios behave in the expected return – risk coordinate system in the function of diversification. We generate 200-200 random equally-weighted portfolios with 2, 5 and 10 securities involved, and compare these to single securities using standard deviation, the CAPM beta, the Shannon entropy and the Rényi entropy as risk measures; the results are presented in Fig. 2.

Fig. 2 shows the performance of random portfolios by diversification using different risk estimation methods. One can see that the characteristics of standard deviation and entropy are quite similar, with the portfolios being situated on a hyperbola as in the portfolio theory of Markowitz [1]. Different characteristics can be observed by using the CAPM beta; the more securities that are involved in a portfolio, the closer they are situated in the center of the coordinate system.

## Long term explanatory power

In order to evaluate how efficiently the risk measures explain the expected risk premium over a long period, we estimate the risk for each security using standard deviation, the CAPM beta, and the Shannon and Rényi entropies based on the full period (denoted by *P1*). The single explanatory variable is the risk measure; the target variable is the expected risk premium of the security. We apply the explanatory power estimation by calculating $\hat{\eta}(\kappa)$ $(R^2)$ for each risk measure.

Fig. 3 shows the efficiency of explaining the expected risk premium by the different risk measures; the expected daily risk premium is presented as a function of risk measure. The CAPM beta performs the worst, with 6.17% efficiency.

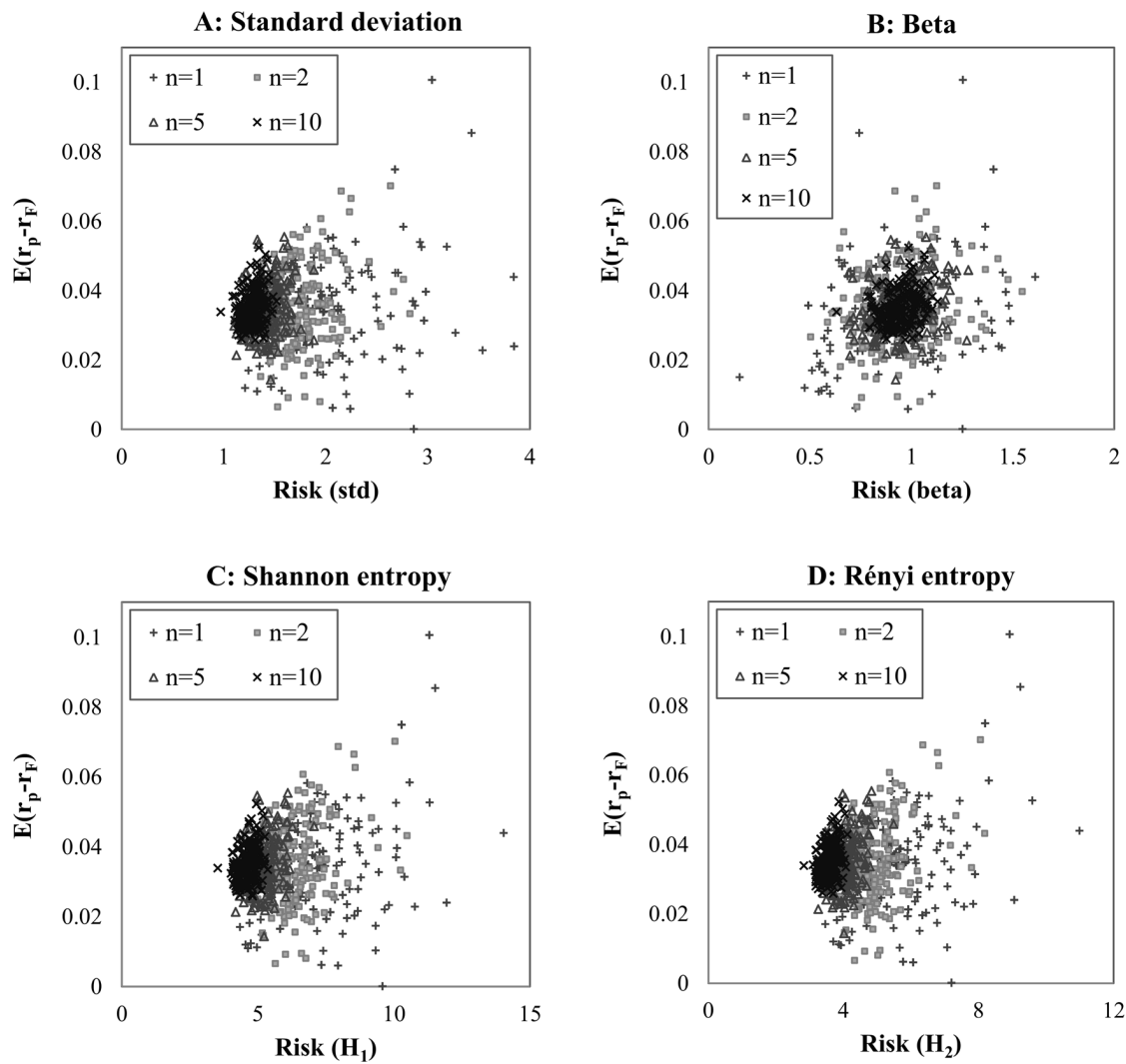

**Fig. 2. Portfolios with different number of securities involved in E(r) – risk system.** *Note:* The panels show the expected risk premium of the portfolios (calculated by the average of daily risk premiums) versus the estimated risk using different methods; the number of securities involved is indicated by the different markers. We generate a sample of 750 random portfolios by using 150 randomly selected securities and 200-200 random equally weighted portfolios with 2, 5 and 10 securities. The risk of portfolios is estimated by standard deviation, CAPM beta, Shannon- and Rényi entropy by using daily returns in the period from 1985 to the end of 2011. Both types of entropy functions are calculated by histogram based density function estimation, with 175 bins for Shannon entropy and 50 bins for Rényi entropy.

doi:10.1371/journal.pone.0115742.g002

However, the explanatory power of standard deviation (7.83%) is higher than that of the CAPM beta, and both entropies perform significantly better, with efficiency of 12.98% for the Shannon entropy and 15.71% for the Rényi entropy. Based on the equation of linear regressions, the average unexplained risk premium (intersect on the *Y*-axis or Jensen alpha [21]) for the entropy methods (0.0091, 0.0059) is lower than that for the standard methods (0.0170 for standard deviation and 0.0209 for the CAPM beta).

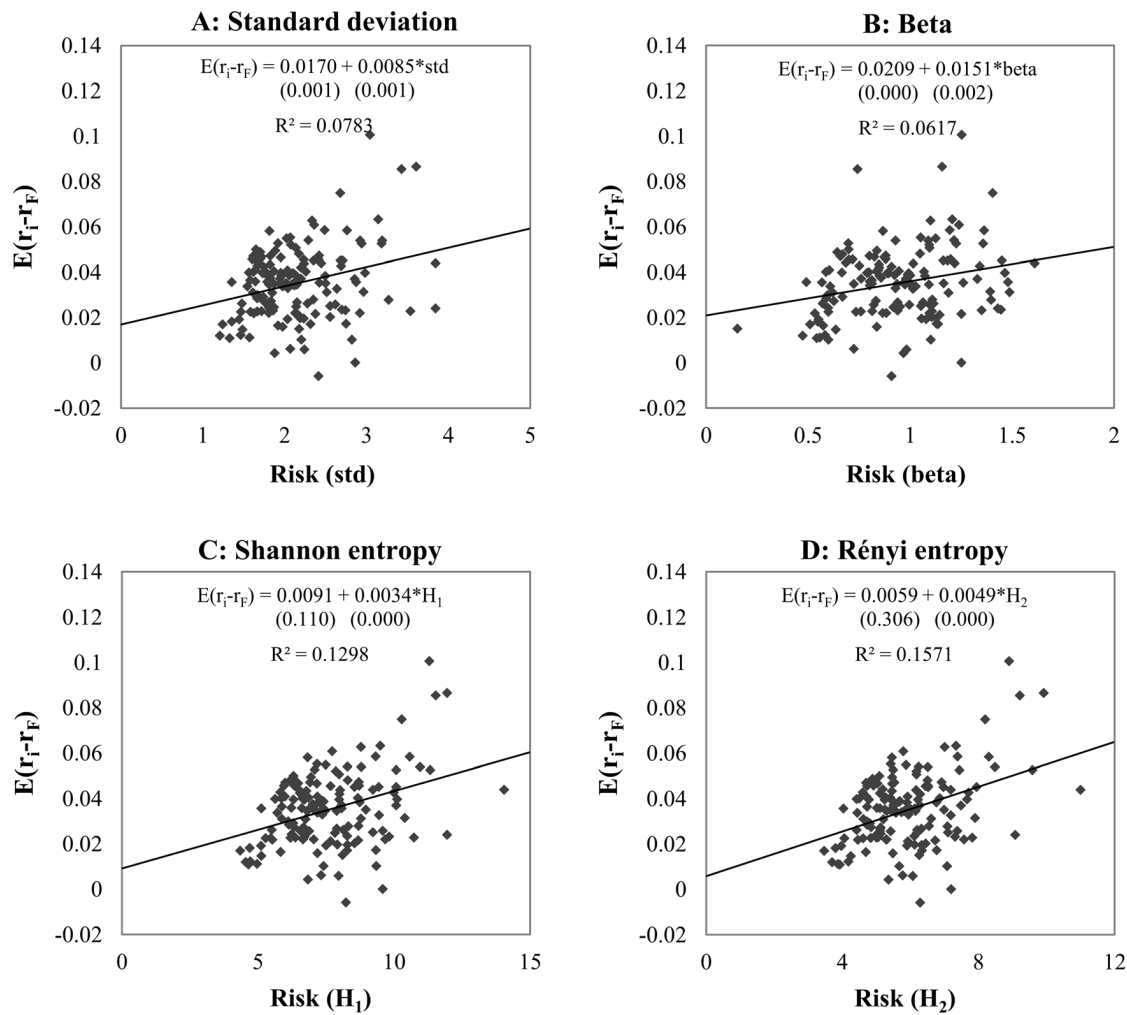

**Fig. 3. Explanatory power of risk measures in long term.** *Note*: The four panels show the relationship between risk premium and risk (standard deviation, CAPM beta, Shannon- and Rényi entropy) of 150 randomly selected securities by using different estimation methods. Both types of entropy functions are calculated by histogram based density function estimation, with 175 bins for Shannon entropy and 50 bins for Rényi entropy. The equation and the explanatory power ($R^2$) of the linear regressions are presented using expected risk premium as target variable and risk as explanatory variable. Under the OLS regression equations in brackets the p-values can be seen for each parameter estimations. The $R^2$s of the models applying entropy based risk measures are significantly different form standard deviation and CAPM beta at 1% level.

doi:10.1371/journal.pone.0115742.g003

We also measure the explanatory power for different numbers of securities involved in the portfolio, by generating at most 100,000 samples for each; we present these results in Fig. 4.

Fig. 4 illustrates how the explanatory power changes with diversification. One can see that the explanatory power of standard deviation and entropy decreases with an increase in the number of securities involved in the portfolio, while the performance of the CAPM beta is nearly constant. While the CAPM beta models the systematic risk only, the standard deviation and entropy are capable of measuring specific risk, which gives additional explanatory power for less-diversified portfolios. Despite the decreased explanatory power of both entropy functions, they perform better than the CAPM beta in all the cases that were investigated. For well-diversified portfolios the explanatory power of the Rényi entropy is 1.5 times higher than that of the CAPM beta.

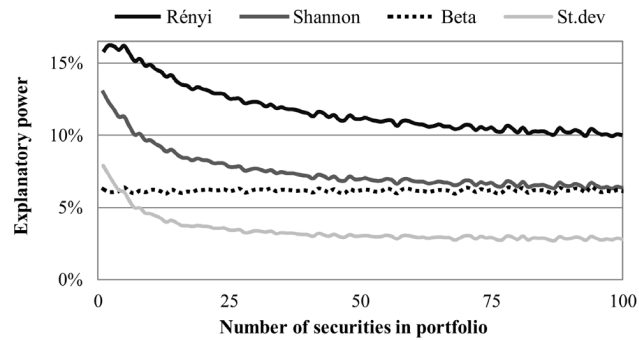

**Fig. 4. Explanatory power of risk measures in long term by diversification.** *Note*: This figure shows the explanatory power ($R^2$) of portfolios with different number of securities involved and different risk estimate methods. We generate 10 million random equally weighted portfolios with different number of securities involved (at most 100,000 for each size) using daily risk premiums of 150 randomly selected securities. The risk of portfolios is estimated by standard deviation (light gray curve), CAPM beta (black dotted curve), Shannon- (gray) and Rényi entropy (black). Both types of entropy functions are calculated by histogram based density function estimation, with 175 bins for Shannon entropy and 50 bins for Rényi entropy.

## Explanatory power by primary market trends

We split the original 27-year sample by primary market trend into a "bullish" and a "bearish" sample (denoted by P1+ and P1-), containing returns for upward and downward periods, respectively (for the labels of the periods see S4 Table). For these two sample sets we investigate the explanatory power for standard deviation, the CAPM beta, and the Shannon and Rényi entropies using the same parameter for the histogram-based entropy estimation as for the previous experiments. Fig. 5 and Fig. 6 show the results in the expected risk premium – risk coordinate system.

Our results for the bullish and bearish regimes show that the different risk measures behave similarly in terms of the positive and negative relationships between risk and return. This behavior underlines the fact that an entropy-based risk measure can give contradictory results in a similar way to traditional risk estimations in different regimes. In bullish market circumstances we find a very high explanatory power for all kinds of risk measures: 33.90%, 36.67%, 43.45% and 42.36% with standard deviation, the CAPM beta, the Shannon entropy and the Rényi entropy, respectively. As for the full sample tests, the slopes of the regression lines are positive, meaning that higher risk-taking promises higher returns. In contrast to the bullish market, during downward trends higher risk-taking does not result in higher returns and, indeed, the higher the risk the higher the negative premium achieved by the investor. We have to mention that the explanatory power of the CAPM beta is higher than that of the entropy-based risk measures. Our entropy results are in line with those for the CAPM beta; and the regime dependency is clear as well. On the other hand, the explanatory power is

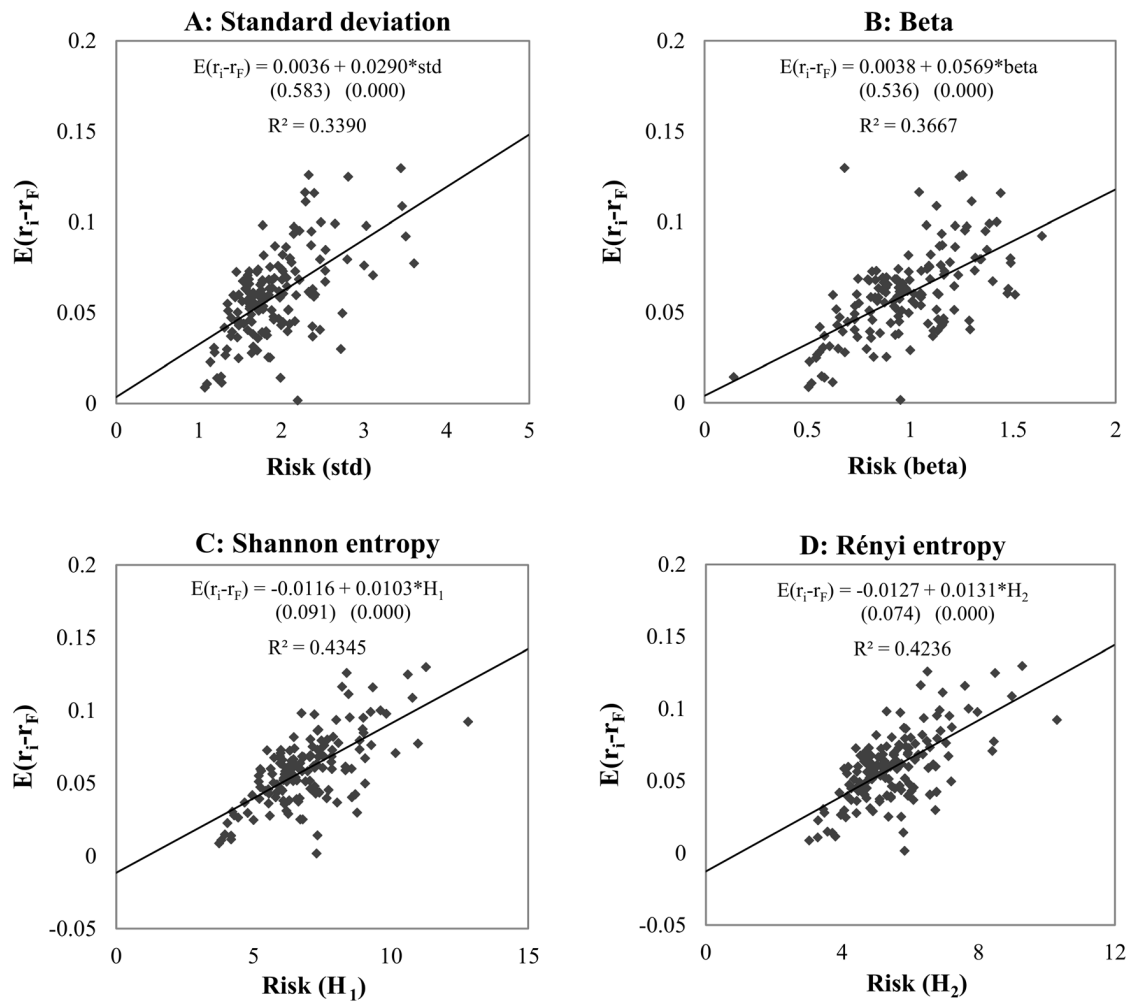

**Fig. 5. Explanatory power of risk measures in bullish sample.** *Note*: The panels show the relationship between the expected risk premium of securities and risk by using different estimation methods. We present the equation of linear regression and the goodness of fit ($R^2$). We estimated the risk of 150 random securities in upward trend periods (bull market) from 1985 to the end of 2011 using standard deviation, CAPM beta, Shannon- and Rényi entropy risk estimation methods. Both types of entropy functions are calculated by histogram based density function estimation, with 175 bins for Shannon entropy and 50 bins for Rényi entropy. Under the OLS regression equations in brackets the p-values can be seen for each parameter estimations. The $R^2$s of the models applying entropy based risk measures are significantly higher than the models with standard deviation and CAPM beta at 1% level.

doi:10.1371/journal.pone.0115742.g005

again much higher for this regime than for the full sample. Altogether, we argue that the test results for the full sample give a better comparison opportunity, as the sample sizes of the bullish and bearish markets are different and at the present moment the investor cannot decide whether there is an upward or a downward trend.

## Short term explanatory and predictive power

Although attractive results are achieved within the sample, this does not necessarily mean high efficiency outside the sample. Therefore we took several ten-year periods, shifting the starting year by one year for each, with the first period being 1985 to 1994 and the last 2002 to 2011. As the full data set covers 27 complete years, we used 18 ten-year periods. We split each ten-year period into two shorter five-year periods (*P2i* and *P2o*), with the risk measures being estimated based on the first period and the predictive efficiency being measured in the second period. In the previous sections, we have presented the results for in-sample for the full sample and for the different regimes, and here we summarize these and we also compare the long-term in-sample results with the short-term in-sample and out-of–sample results.

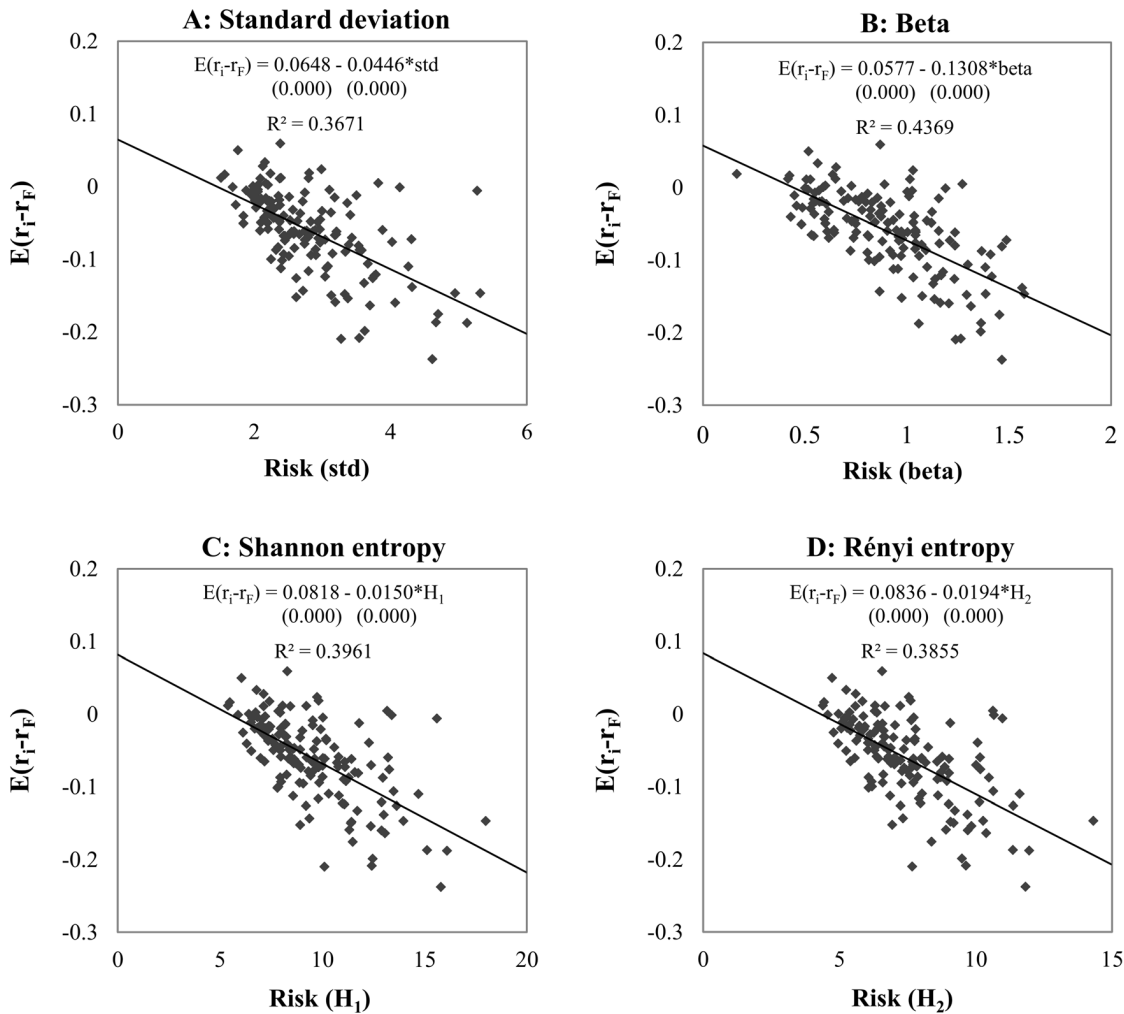

**Fig. 6. Explanatory power of risk measures in bearish sample.** *Note*: The panels show the relationship between the expected risk premium of securities and risk by using different estimation methods. We present the equation of linear regression and the goodness of fit ($R^2$).We estimated the risk of 150 random securities in downward trend periods (bear market) from 1985 to the end of 2011 using standard deviation, CAPM beta, Shannon- and Rényi entropy risk estimation methods. Both types of entropy functions are calculated by histogram based density function estimation, with 175 bins for Shannon entropy and 50 bins for Rényi entropy. Under the OLS regression equations in brackets the p-values can be seen for each parameter estimations. The $R^2$ of the models applying entropy based risk measures are significantly higher than the model with standard deviation at 1% level.

Table 1 summarizes the explanatory power of the investigated risk measures for the different samples. $\hat{\eta}_{P1}, \hat{\eta}_{P1}{}^{+}$, and $\hat{\eta}_{P1}{}^{-}$ show the results of the long-term analysis

**Table 1.** Efficiency of explaining and predicting risk premium in different samples.

| Risk measure | $\bar{\eta}_{P1}$ | $\bar{\eta}_{P1+}$ | $\bar{\eta}_{P1-}$ | $\overline{\bar{\eta}_{P2i}}$ | $\overline{\bar{\eta}_{P2o}}$ | $\sigma_R(\bar{\eta}_{P2i})$ | $\sigma_R(\bar{\eta}_{P2o})$ |
|---|---|---|---|---|---|---|---|
| Standard deviation | 7.83% | 33.9% | 36.7% | 7.94% | 9.7% | 0.75 | 0.65 |
| Beta | 6.17% | 36.7% | 43.7% | 13.31% | 6.45% | 0.98 | 1.02 |
| Shannon entropy | 12.98% | 43.5% | 39.6% | 13.38% | 10.15% | 0.69 | 0.64 |
| Rényi entropy | 15.71% | 42.4% | 38.6% | 12.82% | 9.34% | 0.63 | 0.62 |

*Note:* The table summarizes the explanatory power (in sample $R^2$) of the investigated risk measures in different samples. We estimate risk measures of 150 random securities using standard deviation, CAPM beta, Shannon- and Rényi entropy risk estimation methods for (1) long term, from 1985 to the end of 2011 (1985–2011); (2) long term on upward trends (bull market), (3) long term on downward trends (bear market), (4) 18 10-year periods shifting by one year from period (1985–1994) to period (2002–2011), split into two 5-5 year periods for each. Both types of entropy functions are calculated by histogram based density function estimation, with 175 bins for Shannon entropy and 50 bins for Rényi entropy. The $\bar{\eta}_{P1}$ shows the explanatory power of risk measures for long term, the $\bar{\eta}_{P1+}$, and $\bar{\eta}_{P1-}$ summarizes the explanatory power on upward and downward trends, respectively, $\overline{\bar{\eta}_{P2i}}$ stands for the average explanatory power of risk measured in the first 5 years of 10-year shorter periods in sample, and $\overline{\bar{\eta}_{P2o}}$ shows the average predicting power of risk measures (out of sample $R^2$) calculated by estimating risk in the first 5 years and evaluating them on the other 5 years in each 10-year periods. The last two columns show the relative standard deviation of the explanatory and predicting power based on the 18 shorter periods for the investigated risk measures.

doi:10.1371/journal.pone.0115742.t001

**Table 2.** Explanatory power in short period samples.

| P2i | $\sigma$ | $\beta$ | $H_1$ | $H_2$ | T-test $H_{1,\sigma}$ | T-test $H_{1,\beta}$ | T-test $H_{2,\sigma}$ | T-test $H_{2,\beta}$ | sig $H_1$ ($\sigma$/$\beta$) | sig $H_2$ ($\sigma$/$\beta$) |
|---|---|---|---|---|---|---|---|---|---|---|
| 1985–1989 | 3.2% | 9.4% | 5.1% | 3.7% | 19.3 | −32.8 | 6.7 | −46.0 | ***/ | ***/ |
| 1986–1990 | 1.7% | 3.6% | 2.6% | 3.6% | 11.8 | −12.8 | 22.7 | 0.4 | ***/ | ***/ |
| 1987–1991 | 4.1% | 4.8% | 6.0% | 7.9% | 13.9 | 7.6 | 27.8 | 21.8 | ***/*** | ***/*** |
| 1988–1992 | 5.5% | 5.0% | 6.7% | 6.8% | 7.7 | 8.5 | 8.1 | 8.8 | ***/*** | ***/*** |
| 1989–1993 | 3.5% | 4.2% | 8.9% | 7.2% | 40.6 | 32.9 | 28.9 | 21.4 | ***/*** | ***/*** |
| 1990–1994 | 9.6% | 7.1% | 23.4% | 20.1% | 66.1 | 75.8 | 48.2 | 57.7 | ***/*** | ***/*** |
| 1991–1995 | 16.0% | 13.6% | 28.1% | 21.9% | 67.7 | 56.3 | 32.9 | 30.8 | ***/*** | ***/*** |
| 1992–1996 | 16.3% | 17.8% | 24.4% | 20.5% | 41.8 | 29.1 | 24.4 | 12.4 | ***/*** | ***/*** |
| 1993–1997 | 7.5% | 24.9% | 15.4% | 13.6% | 38.1 | −40.0 | 32.7 | −55.2 | ***/ | ***/ |
| 1994–1998 | 7.0% | 30.1% | 15.8% | 12.6% | 86.1 | −101.1 | 55.6 | −124.3 | ***/ | ***/ |
| 1995–1999 | 16.6% | 51.2% | 28.3% | 27.3% | 88.3 | −166.4 | 77.8 | −167.8 | ***/ | ***/ |
| 1996–2000 | 8.6% | 28.2% | 18.0% | 20.8% | 63.7 | −67.7 | 85.0 | −49.7 | ***/ | ***/ |
| 1997–2001 | 2.1% | 15.3% | 7.2% | 9.7% | 39.3 | −60.0 | 58.2 | −41.7 | ***/ | ***/ |
| 1998–2002 | 0.2% | 2.5% | 1.5% | 2.4% | 28.2 | −16.0 | 41.8 | −0.3 | ***/ | ***/ |
| 1999–2003 | 6.1% | 7.5% | 8.1% | 9.8% | 16.9 | 6.2 | 29.2 | 18.8 | ***/*** | ***/*** |
| 2000–2004 | 1.9% | 0.1% | 1.5% | 1.4% | −8.0 | 35.5 | −9.2 | 34.8 | /*** | /*** |
| 2001–2005 | 15.1% | 5.6% | 17.5% | 18.4% | 16.0 | 93.5 | 22.4 | 102.2 | ***/*** | ***/*** |
| 2002–2006 | 17.9% | 8.9% | 22.3% | 23.1% | 24.6 | 89.6 | 30.2 | 98.8 | ***/*** | ***/*** |
| Average | 7.94% | 13.31% | 13.37% | 12.82% | | | | | | |
| Rel. dev | 0.75 | 0.98 | 0.69 | 0.63 | | | | | | |

*Note:* This table summarizes the explanatory power of the different risk measures for expected risk premium in the first 5 years of 18 10-year periods (P2i) shifting by one year from period (1985–1994) to period (2002–2011). We estimate and evaluate risk measures of 150 randomly selected securities from the S&P500 index using standard deviation ($\sigma$), CAPM beta ($\beta$), Shannon entropy ($H_1$) and Rényi entropy ($H_2$) risk estimation methods by daily risk premiums. Both types of entropy functions are calculated by histogram based density function estimation, with 175 bins for Shannon entropy and 50 bins for Rényi entropy. We apply t-statistics by bootstrapping method to measure whether differences in $R^2$s are significant. We use *s to designate that the entropy based risk measure is significantly higher than the standard deviation and CAPM beta;
***, ** and * stands for 1%, 5% and 10% significance level respectively.

doi:10.1371/journal.pone.0115742.t002

for the full period and during the upward and downward trends, respectively; $\overline{\eta}_{P2i}$, and $\overline{\eta}_{P2o}$ stand for the average efficiency measured for short-term in-sample and out-of-sample, respectively; and $\sigma_R(\hat{\eta}_{P2i})$, and $\sigma_R(\hat{\eta}_{P2o})$ measure the relative standard deviation of the efficiency when applying the in-sample and out-of-sample test for short periods (For the detailed results for all periods see Table 2 and Table 3). While the standard deviation risk measure performs almost the same in the long and the short run (7.83% vs. 7.94), its predictive efficiency is surprisingly good (9.70%). The explanatory power of the CAPM beta in the long period is low (6.17%), while the average efficiency in the short periods is more than twice as high (13.31%). We use arithmetic averages [22]. Comparing the results for in-sample and the out-of-sample, the predictive power of the beta is relatively low (6.45%), which suggests that the model may be over-fitted for the training sample. The Shannon entropy performs better than the standard deviation and the CAPM beta in each sample. The Rényi entropy shows the highest explanatory power in the long run; however, in short periods the Rényi

**Table 3.** Predicting power in short periods out of sample.

| P2i | P2o | $\sigma$ | $\beta$ | $H_1$ | $H_2$ | T-test $H_1,\sigma$ | T-test $H_1,\beta$ | T-test $H_2,\sigma$ | T-test $H_2,\beta$ | sig $H_1$ ($\sigma/\beta$) | sig $H_2$ ($\sigma/\beta$) |
|---|---|---|---|---|---|---|---|---|---|---|---|
| 1985–1989 | 1990–1994 | 7.3% | 2.8% | 13.0% | 10.0% | 35.2 | 80.4 | 15.4 | 52.0 | ***/*** | ***/*** |
| 1986–1990 | 1991–1995 | 17.0% | 4.1% | 19.3% | 18.1% | 12.8 | 101.4 | 5.6 | 85.8 | ***/*** | ***/*** |
| 1987–1991 | 1992–1996 | 21.5% | 5.9% | 22.6% | 17.5% | 7.1 | 99.8 | −23.0 | 71.8 | ***/*** | /*** |
| 1988–1992 | 1993–1997 | 9.8% | 7.9% | 14.6% | 13.2% | 34.9 | 43.9 | 27.8 | 38.1 | ***/*** | ***/*** |
| 1989–1993 | 1994–1998 | 7.9% | 16.5% | 13.5% | 11.6% | 65.8 | −21.3 | 45.5 | −35.8 | ***/ | ***/ |
| 1990–1994 | 1995–1999 | 10.0% | 23.9% | 16.6% | 14.9% | 63.2 | −57.1 | 48.0 | −70.1 | ***/ | ***/ |
| 1991–1995 | 1996–2000 | 9.0% | 14.1% | 9.1% | 9.0% | 0.1 | −44.0 | −0.7 | −45.0 | / | / |
| 1992–1996 | 1997–2001 | 11.3% | 14.7% | 11.7% | 11.8% | 1.7 | −20.8 | 2.0 | −21.0 | */ | **/ |
| 1993–1997 | 1998–2002 | 14.2% | 4.7% | 12.7% | 10.8% | −11.0 | 66.8 | −26.3 | 54.4 | /*** | /*** |
| 1994–1998 | 1999–2003 | 24.7% | 2.7% | 17.5% | 19.8% | −54.6 | 154.7 | −36.3 | 173.0 | /*** | /*** |
| 1995–1999 | 2000–2004 | 3.6% | 6.8% | 0.3% | 0.5% | −59.0 | −90.3 | −55.0 | −87.1 | / | / |
| 1996–2000 | 2001–2005 | 8.0 | 0.0% | 3.6% | 3.0% | −47.1 | 67.9 | −54.5 | 62.7 | /*** | /*** |
| 1997–2001 | 2002–2006 | 10.3% | 0.4% | 6.1% | 4.5% | −38.9 | 91.9 | −56.1 | 78.8 | /*** | /*** |
| 1998–2002 | 2003–2007 | 7.8% | 3.2% | 6.4% | 5.8% | −13.4 | 40.8 | −19.7 | 35.0 | /*** | /*** |
| 1999–2003 | 2004–2008 | 1.5% | 3.2% | 1.9% | 2.1% | 5.5 | −18.5 | 7.7 | −16.3 | ***/ | ***/ |
| 2000–2004 | 2005–2009 | 4.7% | 1.2% | 5.0% | 5.1% | 4.8 | 62.9 | 5.9 | 63.1 | ***/*** | ***/*** |
| 2001–2005 | 2006–2010 | 2.2% | 1.9% | 3.2% | 4.0% | 17.8 | 20.9 | 28.9 | 31.5 | ***/*** | ***/*** |
| 2002–2006 | 2007–2011 | 4.0% | 2.3% | 5.4% | 6.5% | 21.5 | 50.1 | 35.2 | 62.2 | ***/*** | ***/*** |
| **Average** | | **9.70%** | **6.45%** | **10.14%** | **9.34%** | | | | | | |
| **Relative deviation** | | **0.65** | **1.02** | **0.64** | **0.62** | | | | | | |

*Note:* This table summarizes the predicting power of the investigated risk measures for expected risk premium in the last 5 years of 18 10-year periods shifting by one year from period (1985–1994) to period (2002–2011). We estimate risk measures of 150 randomly selected securities from the S&P500 index using standard deviation ($\sigma$), CAPM beta ($\beta$), Shannon entropy ($H_1$) and Rényi entropy ($H_2$) risk estimation methods by daily risk premiums in the first 5 years (P2i) and measure the predicting power on the next 5 years (P2o) by estimating the goodness of fit of linear regression ($R^2$). Both types of entropy functions are calculated by histogram based density function estimation, with 175 bins for Shannon entropy and 50 bins for Rényi entropy. We apply t-statistics by bootstrapping method to measure whether differences in $R^2$s are significant. We use *s to designate that the entropy based risk measure is significantly higher than the standard deviation and CAPM beta; ***, ** and * stands for 1%, 5% and 10% significance level respectively.

doi:10.1371/journal.pone.0115742.t003

entropy performs worse than the Shannon entropy. Comparing the reliability of the risk estimators, the standard deviation of the in-sample and out-of-sample results is the lowest for the entropy risk measures, and the highest for the CAPM beta. Summarizing our results, we state that the beta can beat the entropy only in the case of bearish market circumstances. In any other situation, entropy seems to be a better and more reliable risk measure.

## Conclusions

Entropy as a novel risk measure combines the advantages of the CAPM's risk parameter (beta) and the standard deviation. It captures risk without using any information about the market, and it is capable of measuring the risk reduction effect of diversification. The explanatory power for the expected return within the sample is better than the beta, especially in the long run covering bullish and bearish periods; the predictive power for the expected return is higher than for standard deviation. Both the Shannon and the Rényi entropies give more reliable risk estimation; their explanatory power exhibits significantly lower variance compared to the beta or the standard deviation. If upward and downward trends are distinguished, the regime dependency of entropy can be recognized: this result is similar to that for the beta. Among the entropy estimation methods reviewed, the histogram-based method proved to be the most efficient in terms of explanatory and predictive power; we propose a simple estimation formula for the Shannon and the Rényi entropy functions, which facilitates the application of an entropy-based risk measure.

## Supporting Information

**S1 Table. Descriptive statistics.**
doi:10.1371/journal.pone.0115742.s001 (DOCX)

**S2 Table. The most often used Kernel functions.**
doi:10.1371/journal.pone.0115742.s002 (DOCX)

**S3 Table. Explanatory power of Shannon entropy by different density estimation methods.**
doi:10.1371/journal.pone.0115742.s003 (DOCX)

**S4 Table. Labeling periods by market trend.**
doi:10.1371/journal.pone.0115742.s004 (DOCX)

## Author Contributions

Conceived and designed the experiments: MO DZ. Performed the experiments: MO DZ. Analyzed the data: MO DZ. Contributed reagents/materials/analysis tools: MO DZ. Wrote the paper: MO DZ.

## References

1. **Markowitz H** (1952) Portfolio selection*. The Journal of Finance 7: 77–91. DOI: 10.1111/j.1540-6261.1952.tb01525.x.

2. **Sharpe WF** (1964) Capital asset prices: A theory of market equilibrium under conditions of risk*. The Journal of Finance 19: 425–442. DOI: 10.1111/j.1540-6261.1964.tb02865.x.

3. **Erdős P, Ormos M** (2009) Return calculation methodology: Evidence from the Hungarian mutual fund industry. Acta Oeconomica 59: 391–409. DOI: 10.1556/AOecon.59.2009.4.2.

4. **Erdős P, Ormos M, Zibriczky D** (2011) Non-parametric and semi-parametric asset pricing. Economic Modelling 28: 1150–1162. DOI: 10.1016/j.econmod.2010.12.008.

5. **Clausius R** (1870) XVI. On a mechanical theorem applicable to heat. The London, Edinburgh, and Dublin Philosophical Magazine and Journal of Science 40: 122–127. DOI: 10.1080/147864470 08640370.

6. **Boltzmann L, Hasenöhrl F** (2012) Weitere Studien über das Wärmegleichgewicht unter Gas-molekülen Wissenschaftliche Abhandlungen: Cambridge University Press. DOI: 10.1017/CBO9781139381420.023.

7. **Shannon CE** (1948) A Mathematical Theory of Communication. Bell System Technical Journal 27: 379–423. DOI: 10.1002/j.1538-7305.1948.tb00917.x.

8. **Philippatos GC, Wilson CJ** (1972) Entropy, market risk, and the selection of efficient portfolios. Applied Economics 4: 209–220. DOI: 10.1080/00036847200000017.

9. **Kirchner U, Zunckel C** (2011) Measuring Portfolio Diversification. arXiv preprint arXiv:11024722.

10. **Dionisio A, Menezes R, Mendes DA** (2006) An econophysics approach to analyse uncertainty in financial markets: an application to the Portuguese stock market. The European Physical Journal B 50: 161–164. DOI: 10.1140/epjb/e2006-00113-2.

11. **Maasoumi E, Racine J** (2002) Entropy and predictability of stock market returns. Journal of Econometrics 107: 291–312. DOI: 10.1.1.27.1423.

12. **Nawrocki DN, Harding WH** (1986) State-value weighted entropy as a measure of investment risk. Applied Economics 18: 411–419. DOI: 10.1080/00036848600000038.

13. **Renyi A** (1961) On Measures of Entropy and Information. Fourth Berkeley Symposium on Mathematical Statistics and Probability; Berkeley, Calif. University of California Press. pp. 547–561.

14. **Scott DW** (1979) On optimal and data-based histograms. Biometrika 66: 605–610. DOI: 10.1093/biomet/66.3.605.

15. **Freedman D, Diaconis P** (1981) On the histogram as a density estimator: L2 theory. Probability theory and related fields 57: 453–476. DOI: 10.1007/BF01025868.

16. **Härdle W** (2004) Nonparametric and semiparametric models: Springer. DOI: 10.1007/978-3-642-17146-8.

17. **Silverman BW** (1986) Density estimation for statistics and data analysis: CRC press. Monographs on Statistics and Applied Probability 26. DOI: 10.1007/978-1-4899-3324-9.

18. **Turlach BA** (1993) Bandwidth selection in kernel density estimation: A review: Université catholique de Louvain. DOI: 10.1.1.44.6770.

19. **Beirlant J, Dudewicz EJ, Györfi L, Van der Meulen EC** (1997) Nonparametric entropy estimation: An overview. International Journal of Mathematical and Statistical Sciences 6: 17–40.

20. **Wachowiak MP, Smolikova R, Tourassi GD, Elmaghraby AS** (2005) Estimation of generalized entropies with sample spacing. Pattern Analysis and Applications 8: 95–101. DOI: 10.1007/s10044-005-0247-4.

21. **Jensen MC** (1968) The performance of mutual funds in the period 1945–1964. The Journal of finance 23: 389–416. DOI: 10.1111/j.1540-6261.1968.tb00815.x.

22. **Andor G, Dülk M** (2013) Harmonic mean as an approximation for discounting intraperiod cash flows. The Engineering Economist 58: 3–18. DOI: 10.1080/0013791X.2012.742607.